\pdfoutput=0 

\documentclass[epj]{webofc}
\usepackage[varg]{txfonts}   % Web of Conferences font
\usepackage{array} 
\usepackage{graphicx}
\wocname{\includegraphics[width=0.25cm,clip]{logo} ARENA-2016}
\woctitle{ARENA-2016}
\begin{document}
\title{Tunka-Rex: Status, Plans, and Recent Results}

\author{\firstname{F.~G.} \lastname{Schr\"oder}\inst{1}\fnsep\thanks{\email{frank.schroeder@kit.edu}}
\and
\firstname{P.~A.} \lastname{Bezyazeekov}\inst{2}
\and
\firstname{N.~M.} \lastname{Budnev}\inst{2}
\and
\firstname{O.} \lastname{Fedorov}\inst{2}
\and
\firstname{O.~A.} \lastname{Gress}\inst{2}
\and
\firstname{A.} \lastname{Haungs}\inst{1}
\and
\firstname{R.} \lastname{Hiller}\inst{1}
\and
\firstname{T.} \lastname{Huege}\inst{1}
\and
\firstname{Y.} \lastname{Kazarina}\inst{2}
\and
\firstname{M.} \lastname{Kleifges}\inst{3}
\and
\firstname{E.~E.} \lastname{Korosteleva}\inst{4}
\and
\firstname{D.} \lastname{Kostunin}\inst{1}
\and
\firstname{O.} \lastname{Kr\"omer}\inst{3}
\and
\firstname{V.} \lastname{Kungel}\inst{1}
\and
\firstname{L.~A.} \lastname{Kuzmichev}\inst{4}
\and
\firstname{N.} \lastname{Lubsandorzhiev}\inst{4}
\and
\firstname{R.~R.} \lastname{Mirgazov}\inst{2}
\and
\firstname{R.} \lastname{Monkhoev}\inst{2}
\and
\firstname{E.~A.} \lastname{Osipova}\inst{4}
\and
\firstname{A.} \lastname{Pakhorukov}\inst{2}
\and
\firstname{L.} \lastname{Pankov}\inst{2}
\and
\firstname{V.~V.} \lastname{Prosin}\inst{4}
\and
\firstname{G.~I.} \lastname{Rubtsov}\inst{5}
\and
%\firstname{F.~G.} \lastname{Schr\"oder}\inst{1}
%\and
\firstname{R.} \lastname{Wischnewski}\inst{6}
\and
\firstname{A.} \lastname{Zagorodnikov}\inst{2}
~(Tunka-Rex Collaboration) 
}

\institute{
Institut f\"ur Kernphysik, Karlsruhe Institute of Technology (KIT), Karlsruhe, Germany  
\and
Institute of Applied Physics, Irkutsk State University (ISU), Irkutsk, Russia  
\and
Institut f\"ur Prozessdatenverarbeitung und Elektronik, Karlsruhe Institute of Technology (KIT), Germany
\and
Skobeltsyn Institute of Nuclear Physics, Lomonossov University (MSU), Moscow, Russia
\and
Institute for Nuclear Research of the Russian Academy of Sciences, Moscow, Russia  
\and
Deutsches Elektronen-Synchrotron (DESY), Zeuthen, Germany
}

\abstract{
Tunka-Rex, the Tunka Radio extension at the TAIGA facility (Tunka Advanced Instrument for cosmic ray physics and Gamma Astronomy) in Siberia, has recently been expanded to a total number of 63 SALLA antennas, most of them distributed on an area of one square kilometer.
In the first years of operation, Tunka-Rex was solely triggered by the co-located air-Cherenkov array Tunka-133. 
The correlation of the measurements by both detectors has provided direct experimental proof that radio arrays can measure the position of the shower maximum. 
The precision achieved so far is $40\,$g/cm$^2$, and several methodical improvements are under study. 
Moreover, the cross-comparison of Tunka-Rex and Tunka-133 shows that the energy reconstruction of Tunka-Rex is precise to $15\,\%$, with a total accuracy of $20\,\%$ including the absolute energy scale. 
By using exactly the same calibration source for Tunka-Rex and LOPES, the energy scale of their host experiments, Tunka-133 and KASCADE-Grande, respectively, can be compared even more accurately with a remaining uncertainty of about $10\,\%$. 
The main goal of Tunka-Rex for the next years is a study of the cosmic-ray mass composition in the energy range above $100\,$PeV: 
For this purpose, Tunka-Rex now is triggered also during daytime by the particle detector array Tunka-Grande featuring surface and underground scintillators for electron and muon detection.
}
\maketitle
\section{Introduction}
\label{sec_intro}
Tunka-Rex started in 2012 as radio extension of the air-Cherenkov array Tunka-133 in Siberia close to lake Baikal~\cite{Tunka133_NIM2014}. 
Its original goals have been a cross-calibration of the radio and the air-Cherenkov signal~\cite{TunkaRex_Xmax2016}, and the demonstration that radio detection can indeed by economic compared to other techniques - at least when not operated as stand-alone detector, but as add on to an existing array~\cite{TunkaRex_NIM_2015}. 
After both original goals have been fulfilled within the first three years of operation, the new primary goal of Tunka-Rex is the determination of the mass-composition in the energy range of $10^{17}-10^{18}$ where a transition of cosmic rays from galactic to extra-galactic origin is assumed~\cite{2013ApelKG_LightAnkle}. 
For this goal, the reconstruction methods for the atmospheric depth of the shower maximum, $X_\mathrm{max}$, shall be improved, and the radio measurements of Tunka-Rex shall be analyzed in combination with the co-located Tunka-Grande array of on- and under-ground scintillator detectors~\cite{TAIGA_2014}. 
A secondary goal of Tunka-Rex remains the further investigation of the radio signal, e.g., by comparing the predictions of Monte Carlo simulations codes to measurements, and by testing results and methods found by other experiments.

\begin{figure*}[t]
  \centering
  \includegraphics[width=0.7\linewidth]{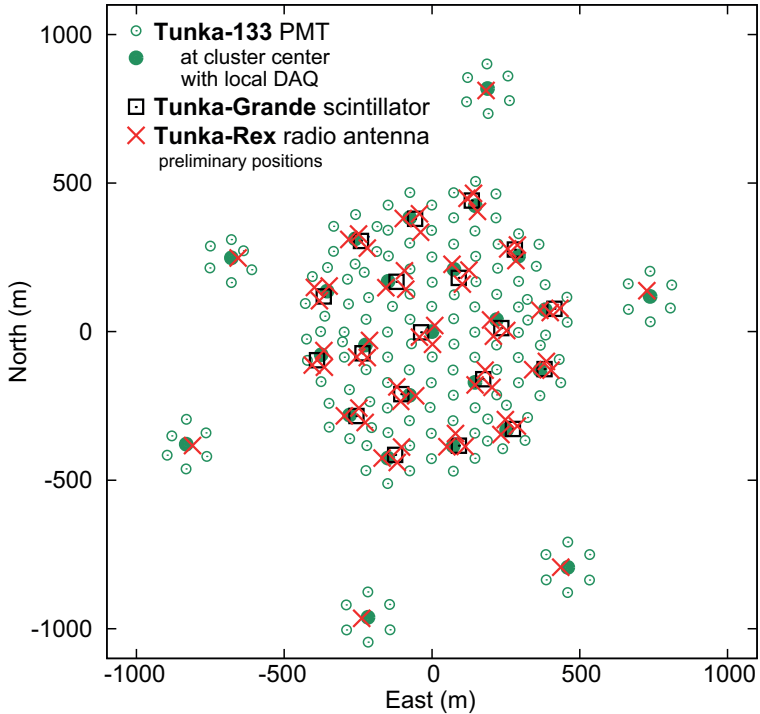}
  \caption{Map of the antenna array Tunka-Rex, the photomultiplier (PMT) array Tunka-133, and the scintillator array Tunka-Grande.}
  \label{fig_tunkaRexMap}
\end{figure*}

\section{Experimental setup}
Since this summer Tunka-Rex consists of 63 antenna stations in a hexagonal structure, each consisting of two orthogonal SALLA antennas measuring the radio emission between $30$ and $80\,$MHz with an effective band of $35-76\,$MHz~\cite{AERAantennaPaper2012}. 
Apart from six outer satellite stations, the antennas are clustered in triplets with antenna spacings of several $10\,$m inside of a triplet and about $200\,$m between the triplets (see figure \ref{fig_tunkaRexMap}). 
This structure is unprecedented and future analyses will show whether it lowers the detection threshold and increases the reconstruction accuracy for $X_\mathrm{max}$ compared to the first years with only one antenna per cluster (see table \ref{tab_tunkaHistory}).
The reason for the new non-uniform layout is mostly technical. 
Each Tunka-133 cluster consists of 7 photomultiplier stations for the detection of air-Cherenkov light and features its own local data-acquisition (DAQ) in the center. 
The first 25 Tunka-Rex antenna stations are attached to this DAQ which provides the trigger during clear nights. 
Moreover, close to each of the 19 inner cluster centers of Tunka-133 there is one station of the particle-detector array Tunka-Grande consisting of scintillators from the dismantled KASCADE-Grande experiment~\cite{Apel2010KASCADEGrande}. 
Meanwhile two antennas per cluster are attached to the DAQ of Tunka-Grande, triggering Tunka-Rex during daytime and bad weather. 
Despite the different local DAQs, the configuration is such that for all events all three antennas of a cluster are read out, i.e., no information is lost. 

The antennas have been calibrated with the same external reference source used by LOPES~\cite{2015ApelLOPES_improvedCalibration} and LOFAR~\cite{NellesLOFAR_calibration2015} with a total accuracy of $22\,\%$ dominated by a $17\,\%$ uncertainty of the absolute scale~\cite{TunkaRex_NIM_2015}. 
Because a large part of the scale uncertainty is correlated between the experiments, their results can be compared to each other with a relative uncertainty of only $10\,\%$, which enables more precise cross-checks of results by the different experiments~\cite{TunkaRexScale2016}. 
Recently, we also studied the timing accuracy in more detail, since a relative accuracy as good as a nanosecond would be required to exploit timing information for $X_\mathrm{max}$ reconstruction~\cite{2014ApelLOPES_wavefront}. 
A previous analysis with a constant-wave reference beacon similar to the ones used at LOPES~\cite{SchroederTimeCalibration2010} or AERA~\cite{AERAairplanePaper2015} revealed that the relative timing is indeed stable to better than a nanosecond during single nights~\cite{SchroederTunkaRex_PISA2015}. 
However, due to daily recalibration of the clock system, there are jumps between nights of several nanoseconds, and moreover the offsets between different clusters are only known to $5\,$ns accuracy. 
Furthermore, background (= internal + external noise) contributes significantly to the uncertainty of measured pulse times, such that both, the synchronization and the noise uncertainty, are comparable in size to the residuals between the real wavefront and the shower plane (see table \ref{tab_timing}). 
This situation is sufficient for a $1-2^\circ$ resolution on the arrival direction, but significant efforts are still necessary to make the timing of Tunka-Rex accurate enough for a measurement of the radio wavefront. 
Thus, the current standard methods for reconstruction of the shower energy and of $X_\mathrm{max}$ rely fully on amplitude measurements ~\cite{KostuninTheory2015, Kostunin_ARENA2016}, but do not yet exploit other observables like the pulse time.

\begin{table}[t]
\centering
\begin{minipage}[t]{0.55\linewidth}
\caption{Stages of the Tunka-Rex experiment. Measurements take place in \lq seasons\rq~from autumn to spring and are interrupted each summer for several reasons.} \label{tab_tunkaHistory}
%\vspace{0.3cm}
\small
\begin{tabular}{lccc}
\hline
Season&Number & \multicolumn{2}{c}{trigger by Tunka} \\
&of antennas& -133 & -Grande\\
\hline
Oct.~2012 - Apr.~2013 & 18 & x & \\
Sep.~2013 - Mar.~2014 & 25 & x & \\
Oct.~2014 - Apr.~2015 & 25 & * & \\
Sep.~2015 - Jun.~2016 & 44 & x & x\\
from autumn 2016 on  & 63 & x & x\\
\hline
\multicolumn{3}{l}{* \footnotesize Data not used due to timing problem.}
\end{tabular}
\end{minipage}
\hfill
\begin{minipage}[t]{0.4\linewidth}
\caption{Average contributions to the time uncertainty of Tunka-Rex studied by comparing real events with CoREAS simulations before and after adding background to the simulations.} \label{tab_timing}
\small
\begin{tabular}{lc}
\hline
Source of uncertainty & size (ns) \\
\hline
Synchronization & 6 \\
Noise & 5 \\
Plane wavefront model & 5 \\
\textbf{total (squared sum)} & \textbf{9}\\
\hline
\end{tabular}
\end{minipage}
\end{table}

\section{Analysis results}
In the first two seasons 178 events have been measured with sufficient signal-to-noise ratio in at least 3 antennas, and with the arrival directions reconstructed by Tunka-133 and Tunka-Rex in agreement~\cite{TunkaRex_Xmax2016}. 
For about half of the events, the absolute amplitude has been compared to CoREAS simulations~\cite{TunkaRex_NIM_2015}, and LOPES measurements~\cite{TunkaRexScale2016}. 
In both cases, Tunka-Rex measurements are in agreement within respective uncertainties of about $20\,\%$ and $10\,\%$. 
Furthermore, the primary energy and $X_\mathrm{max}$ have been reconstructed from the radio measurements and compared to the air-Cherenkov measurements, again, finding agreement. 
The derived precision for the radio measurement is about $15\,\%$ for the shower energy, and about $40\,$g/cm$^2$ for $X_\mathrm{max}$, the latter though only for a subset of high-quality events~\cite{TunkaRex_Xmax2016}. 
As shown at this conference, the shower energy can even be reconstructed from a single amplitude measurements at one antenna station, when using the shower axis provided by the host experiment Tunka-133~\cite{Hiller_ARENA2016}. 
While the accuracy is slightly worse in this case, this lowers the energy threshold and might be an interesting option for future hybrid analyses combining Tunka-Rex with muon measurements by Tunka-Grande~\cite{SchroederReview2016}.

\section{Conclusion}
Tunka-Rex has shown that a sparse and cost-effective radio array can compete with more expensive arrays in several aspects, in particular with respect to the precision of the shower energy. 
For $X_\mathrm{max}$, Tunka-Rex has brought experimental evidence that the radio technique can indeed be used to measure the distance to the shower maximum, complementing earlier evidence by LOPES~\cite{2012ApelLOPES_MTD}.
However, the $X_\mathrm{max}$ of Tunka-Rex is still two times worse than that of optical detection methods or of the dense radio array LOFAR~\cite{BuitinkLOFAR_Xmax2014}, but we expect that the recently increased density of the array as well as the optimization of our reconstruction methods will improve the precision of Tunka-Rex. 
In addition, the combination of radio measurements with the electron and muon measurements by Tunka-Grade ought to increase the total accuracy on the mass composition for the energy range about $100\,$PeV.

\section*{Acknowledgements}
The construction of Tunka-Rex was funded by the German Helmholtz association and the Russian Foundation for Basic Research (grant HRJRG-303). 
Moreover, this work has been supported by the Helmholtz Alliance for Astroparticle Physics (HAP), by Deutsche Forschungsgemeinschaft (DFG) grant SCHR 1480/1-1, and by the Russian grant RSF 15-12-20022. 

\bibliography{arena2016.bib}

\begin{thebibliography}{20}

\bibitem{Tunka133_NIM2014}
{V.V.~Prosin, et al.~- Tunka-133 Coll.}, Nucl. Inst. Meth. A \textbf{756}, 94
  (2014)

\bibitem{TunkaRex_Xmax2016}
{P.A.~Bezyazeekov, et al.~- Tunka-Rex Coll.}, JCAP \textbf{01}, 052 (2016)

\bibitem{TunkaRex_NIM_2015}
{P.A.~Bezyazeekov, et al.~- Tunka-Rex Coll.}, Nucl. Inst. Meth. A \textbf{802},
  89 (2015)

\bibitem{2013ApelKG_LightAnkle}
{W.D.~Apel, et al.~- KASCADE-Grande Coll.}, Physical Review D \textbf{87},
  081101(R) (2013)

\bibitem{TAIGA_2014}
{N.~M. Budnev, et al.~- TAIGA Coll.}, JINST \textbf{9}, C09021 (2014)

\bibitem{AERAantennaPaper2012}
{Pierre Auger Coll. et al.}, JINST \textbf{7}, P10011 (2012)

\bibitem{Apel2010KASCADEGrande}
{W.D.~Apel, et al.~- KASCADE-Grande Coll.}, Nucl. Inst. Meth. A \textbf{620},
  202 (2010)

\bibitem{2015ApelLOPES_improvedCalibration}
{W.D.~Apel, et al.~- LOPES Coll.}, Astropart. Phys. \textbf{75}, 72 (2016)

\bibitem{NellesLOFAR_calibration2015}
{A.~Nelles et al.}, JINST \textbf{10}, P11005 (2015)

\bibitem{TunkaRexScale2016}
{W.D.~Apel, et al.~- LOPES and Tunka-Rex Colls.}, Physics Letters B
  \textbf{763}, 179 (2016)

\bibitem{2014ApelLOPES_wavefront}
{W.D.~Apel, et al.~- LOPES Coll.}, JCAP \textbf{09}, 025 (2014)

\bibitem{SchroederTimeCalibration2010}
{F.G.~Schr\"oder} et~al., Nucl. Inst. Meth. A \textbf{615}, 277 (2010)

\bibitem{AERAairplanePaper2015}
{Pierre Auger Coll. et al.}, JINST \textbf{11}, P01018 (2016)

\bibitem{SchroederTunkaRex_PISA2015}
{F.G.~Schr\"oder, et al.~Tunka-Rex Coll.}, Nucl. Inst. Meth. A \textbf{824},
  652 (2016)

\bibitem{KostuninTheory2015}
D.~{Kostunin} et~al., Astropart. Phys. \textbf{74}, 79 (2016)

\bibitem{Kostunin_ARENA2016}
{D. Kostunin, et al.~- Tunka-Rex Coll.}, EPJ Web of Conf. \textbf{this issue},
  arXiv:1611.09127 (2016)

\bibitem{Hiller_ARENA2016}
{R. Hiller, et al.~- Tunka-Rex Coll.}, EPJ Web of Conf. \textbf{this issue},
  arXiv:1611.09614 (2016)

\bibitem{SchroederReview2016}
{F.G.~Schr\"oder}, Prog. Part. Nucl. Phys. \textbf{93}, 1 (2017),
  arXiv:1607.08781

\bibitem{2012ApelLOPES_MTD}
{W.D.~Apel, et al.~- LOPES Coll.}, Physical Review D \textbf{85}, 071101(R)
  (2012)

\bibitem{BuitinkLOFAR_Xmax2014}
{S.~Buitink et al.~-LOFAR Coll.}, Phys. Rev. D \textbf{90}, 082003 (2014)

\end{thebibliography}

\end{document}